\documentclass[superscriptaddress, aps, prb, twocolumn, showpacs]{revtex4}

\usepackage[pdftex]{graphicx}
\usepackage{amsmath}

\begin{document}

\title{Nonlinear and time-resolved optical study of the 112-type iron-based superconductor parent Ca$_{1-x}$La$_{x}$FeAs$_{2}$ across its structural phase transition}

\author{J. W. Harter}
\affiliation{Department of Physics, California Institute of Technology, Pasadena, California 91125, USA}
\affiliation{Institute for Quantum Information and Matter, California Institute of Technology, Pasadena, California 91125, USA}

\author{H. Chu}
\affiliation{Institute for Quantum Information and Matter, California Institute of Technology, Pasadena, California 91125, USA}
\affiliation{Department of Applied Physics, California Institute of Technology, Pasadena, California 91125, USA}

\author{S. Jiang}
\affiliation{Department of Physics and Astronomy and California NanoSystems Institute, University of California, Los Angeles, California 90095, USA}

\author{N. Ni}
\affiliation{Department of Physics and Astronomy and California NanoSystems Institute, University of California, Los Angeles, California 90095, USA}

\author{D. Hsieh}
\email[Author to whom correspondence should be addressed. Email address: ]{dhsieh@caltech.edu}
\affiliation{Department of Physics, California Institute of Technology, Pasadena, California 91125, USA}
\affiliation{Institute for Quantum Information and Matter, California Institute of Technology, Pasadena, California 91125, USA}

\date{\today}

\begin{abstract}
The newly discovered 112-type ferropnictide superconductors contain chains of As atoms that break the tetragonal symmetry between the $a$ and $b$ axes. This feature eliminates the need for uniaxial strain that is usually required to stabilize large single domains in the electronic nematic state that exists in the vicinity of magnetic order in the iron-based superconductors. We report detailed structural symmetry measurements of 112-type Ca$_{0.73}$La$_{0.27}$FeAs$_{2}$ using rotational anisotropy optical second harmonic generation. This technique is complementary to diffraction experiments and enables a precise determination of the point group symmetry of a crystal. By combining our measurements with density functional theory calculations, we uncover a strong optical second harmonic response of bulk electric dipole origin from the Fe and Ca $3d$-derived states that enables us to assign $C_2$ as the crystallographic point group. This makes the 112-type materials high-temperature superconductors without a center of inversion, allowing for the possible mixing of singlet and triplet Cooper pairs in the superconducting state. We also perform pump-probe transient reflectivity experiments that reveal a 4.6 THz phonon mode associated with the out-of-plane motion of As atoms in the FeAs layers. We do not observe any suppression of the optical second harmonic response or shift in the phonon frequency upon cooling through the reported monoclinic-to-triclinic transition at 58 K. This allows us to identify $C_1$ as the low-temperature crystallographic point group but suggests that structural changes induced by long-range magnetic order are subtle and do not significantly affect electronic states near the Fermi level.
\end{abstract}


\maketitle

\section{Introduction}

The iron-based superconductors feature an antiferromagnetically ordered phase at low chemical doping that evolves into a superconducting phase upon further doping. In most cases, such as in the 1111- and 122-type materials, the long-range antiferromagnetic order is preceded in temperature by a structural transition that lowers the lattice symmetry from tetragonal to orthorhombic \cite{lumsden2010,paglione2010,stewart2011}. Theoretical and experimental evidence suggests that this intermediate ``nematic'' phase is induced by incipient spin fluctuations above the magnetic ordering temperature \cite{fang2008,nandi2010,fernandes2010}. The electronic nematic state within this symmetry-broken phase is manifested through an array of striking behaviors, including a large in-plane resistivity anisotropy \cite{chu2010} and a substantial splitting of orthogonal energy bands \cite{yi2011}, with important ramifications for superconductivity in this class of materials. In compounds studied thus far, spatially-averaged signatures of nematic symmetry breaking are obscured by the presence of orthorhombic domains, compelling researchers to detwin samples via uniaxial strain. How the strain itself contributes to properties measured this way is also an active area of research \cite{man2015}.

Recently, a new family of iron-based superconductors---the 112-type materials---have been discovered. These materials contain zigzag As chains that break the tetragonal symmetry between the $a$ and $b$ axes, making samples intrinsically detwinned without the application of uniaxial strain. In addition, in contrast to all other known high-temperature superconductors, the reported crystal structure of the 112-type materials lacks a center of inversion \cite{katayama2013}. This is noteworthy because noncentrosymmetric superconductors in the presence of spin-orbit coupling can show a mixing of spin-singlet and spin-triplet Cooper pairs \cite{yip2014}, and spin-orbit coupling is expected to be non-negligible in the 112-type materials because of the presence of heavy rare earth atoms.

\begin{figure*}[htb]
\includegraphics{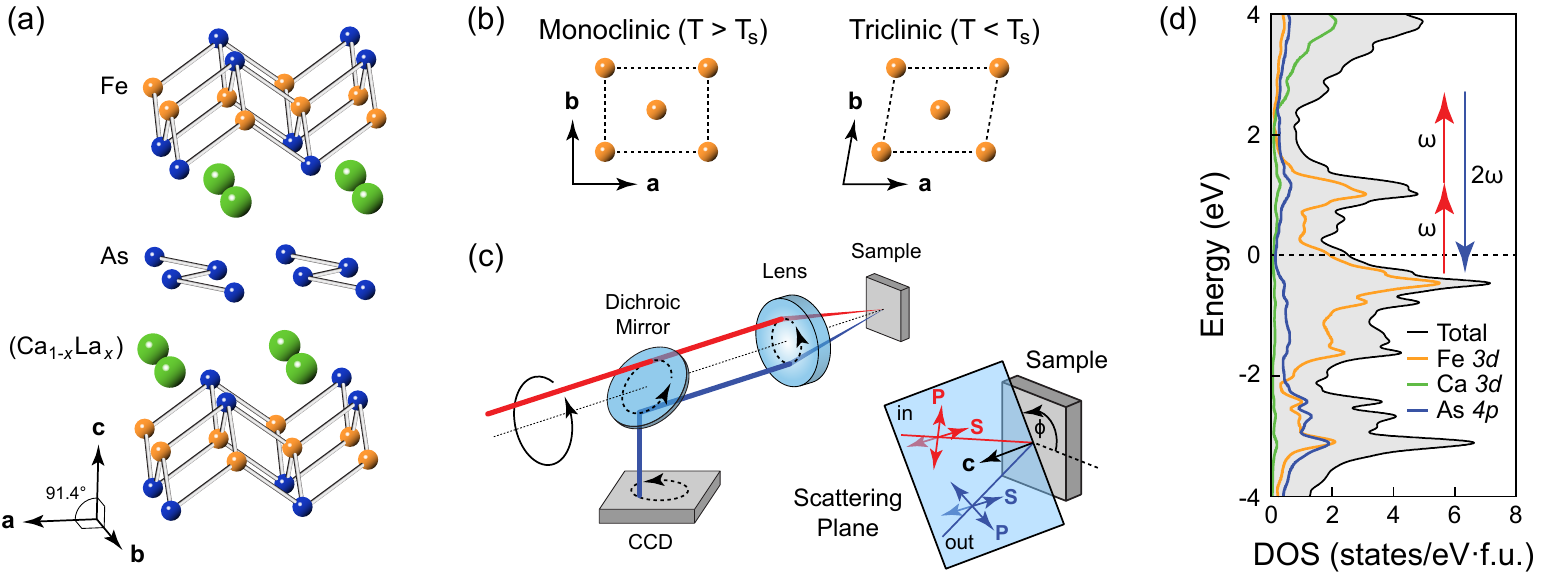}
\caption{\label{figure0} Structure of Ca$_{1-x}$La$_{x}$FeAs$_{2}$ and experimental setup. (a) Crystal structure of Ca$_{1-x}$La$_{x}$FeAs$_{2}$ in the monoclinic phase ($T > T_S$). Zigzag As chains run along the $b$-axis. The space group of the crystal, $P2_1$ (point group $C_2$), is noncentrosymmetric and possesses a single symmetry element: a two-fold screw axis along \textbf{b}. (b) Iron atomic planes above and below the structural transition temperature $T_S$ (distortions from a perfect square lattice are exaggerated for clarity). Below $T_S$ the crystal loses its screw axis and becomes triclinic. (c) Simplified schematic of the RA-SHG experimental setup. A laser beam (800 nm; red), incident from the left, passes through a dichroic mirror and is focused by a lens onto the sample. The reflected second harmonic (400 nm; blue) is recollimated by the lens and reflects off the mirror into a charge-coupled device (CCD) detector. Additional optics used for polarization compensation are omitted for clarity but are described in Ref.~\citenum{harter2015}. By rotating the position of the incident laser beam about the central optical axis, the scattering plane rotates by an angle $\phi$ about the sample surface normal, simultaneously tracing out a circle on the CCD. Bottom right inset shows linear polarizations in (P) and out of (S) the scattering plane. (d) Electrical resistivity (top) and magnetic susceptibility (bottom) measurements showing clear anomalies at the structural and magnetic transition temperatures, respectively. (e) Partial and total density of states of CaFeAs$_2$ as computed by DFT. States near the Fermi level have predominantly Fe $3d$ character, mixing with Ca $3d$ orbitals well above $E_F$. Vertical arrows show the relative scale of the electronic energy bands and the fundamental ($\omega$) and second harmonic ($2\omega$) photons.}
\end{figure*}

The 112-type materials have chemical formula Ca$_{1-x}$\textit{RE}$_{x}$FeAs$_{2}$, where \textit{RE} is a rare earth atom \cite{katayama2013,yakita2014,sala2014}. In these compounds, substitution of Ca by rare earth ions is necessary for structural stability; the $x = 0$ endmember is unstable and cannot be synthesized. One study of \textit{RE} = La has reported an enhancement of antiferromagnetism with increasing $x$, with a maximum N\'{e}el transition temperature at $x > 0.24$ \cite{kawasaki2015}. In addition, the highest superconducting critical temperatures, exceeding 40 K, have only been achieved via a secondary doping mechanism such as chemical substitution or off-stoichiometry \cite{kudo2014,zhou2014}. For these reasons, it is argued that Ca$_{1-x}$La$_{x}$FeAs$_{2}$ with $x \approx 0.27$ should be thought of as the ``parent'' of the family, with superconductivity induced by hole or electron doping away from this parent state \cite{jiang2015}. As illustrated in Figs.~\ref{figure0}(a) and \ref{figure0}(b), the crystal structure of Ca$_{0.73}$La$_{0.27}$FeAs$_{2}$ is unique among the wider family of iron-based superconductors. Instead of showing a tetragonal-to-orthorhombic structural phase transition, the material undergoes a monoclinic-to-triclinic transition at $T_S = 58$ K followed closely by long range antiferromagnetic order at $T_N = 54$ K \cite{jiang2015}. This unusually low crystalline symmetry is due to the presence of zigzag As chains running along the $b$-axis, breaking the symmetry between the $a$ and $b$ axes and distorting the atoms in the unit cell into a noncentrosymmetric arrangement. How the 112-type materials compare with the other iron-based superconductors---and in particular whether or not an electronic nematic state exists in the intermediate region between $T_S$ and $T_N$---is an open question. In attempting to answer such a question, it is imperative to understand the as-yet unknown detailed point group symmetry changes that occur at the reported monoclinic-to-triclinic transition.

When employed to determine crystallographic structure, diffraction experiments generally require accurate measurements of a multitude of Bragg reflections and the analysis of such data through nontrivial structural refinement routines. It can therefore be challenging to detect minute changes in symmetry from such techniques. Rotational anisotropy second harmonic generation (RA-SHG) is an alternative method that relies on carefully measuring the tensor optical response of a crystal in order to determine the crystallographic and electronic point group symmetries of solids \cite{torchinsky2014,torchinsky2015}. As illustrated in Fig.~\ref{figure0}(c), in this technique an incident laser beam is focused onto a sample and the intensity of the second harmonic of the reflected light is measured as the angle $\phi$ between the scattering plane and a well-defined direction parallel to the sample surface is swept. By selecting a combination of different incoming and outgoing light polarizations---typically linear polarized either in (P) or out of (S) the scattering plane---the nonzero elements of the nonlinear optical susceptibility tensor $\chi$ can be deduced from the measured rotational anisotropy. In the electric dipole approximation of SHG, $\chi$ is a third-rank tensor that relates the incident electric field components at frequency $\omega$ to the radiating dipole response $P$ of the material at frequency $2\omega$ \cite{boyd1991}:
\begin{equation}
\label{eqn1}
P_i(2\omega) \propto \chi_{ijk}E_j(\omega)E_k(\omega).
\end{equation}
By Neumann's principle, the symmetries of the crystal completely determine the susceptibility tensor elements that are independent and nonzero. As a consequence, the underlying crystallographic point group symmetries of the sample under study can be deduced by carefully measuring the susceptibility tensor. Accordingly, one can in principle study the symmetry changes across the monoclinic-to-triclinic transition in Ca$_{1-x}$La$_{x}$FeAs$_{2}$ using RA-SHG.

\begin{figure*}[htb]
\includegraphics{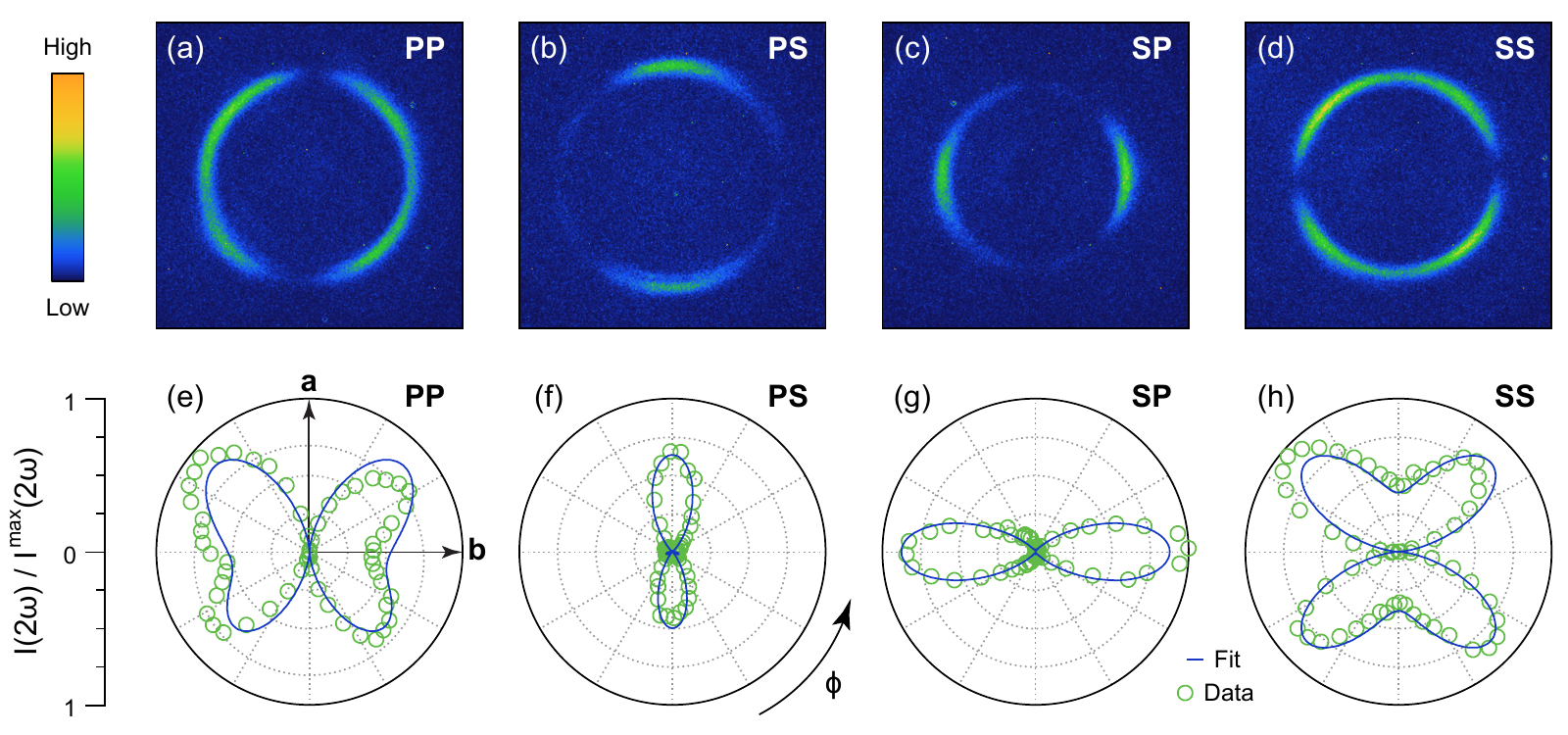}
\caption{\label{figure1} RA-SHG data for Ca$_{0.73}$La$_{0.27}$FeAs$_{2}$ taken at 70 K. Raw CCD images for (a) P$_\mathrm{in}$-P$_\mathrm{out}$, (b) P$_\mathrm{in}$-S$_\mathrm{out}$, (c) S$_\mathrm{in}$-P$_\mathrm{out}$, and (d) S$_\mathrm{in}$-S$_\mathrm{out}$ polarization geometries. (e)-(h) Corresponding rotational anisotropy polar plots extracted from the CCD images. A single intensity scale is maintained across the figures and relative intensities are accurately depicted. Solid curves show a global fit to the data assuming a $C_2$ point group, as discussed in the text.}
\end{figure*}

\section{Experimental Details}

Details concerning the growth and characterization of Ca$_{1-x}$La$_{x}$FeAs$_{2}$ single crystals can be found in Ref.~\citenum{jiang2015}. As Fig.~\ref{figure0}(d) shows, resistivity and magnetic susceptibility measurements on the batch of samples used in our study are consistent with those reported in Ref.~\citenum{jiang2015}. To assess the orbital character of the low-energy electronic states, density functional theory (DFT) calculations were performed on the hypothetical compound CaFeAs$_2$ using the WIEN2k software package \cite{blaha2001} with the unit cell parameters of Ref.~\citenum{katayama2013}. Treating La substitution within a rigid band picture in which one free electron per La atom is added to the system, the Fermi level ($E_F$) will shift by a negligible amount ($\sim 0.1$ eV) for $x \sim 0.27$. RA-SHG experiments were performed on the (001) crystal cleavage plane. Samples were cleaved in air and then immediately placed in vacuum (pressure below $10^{-7}$ Torr) before measurement. We utilized a RA-SHG setup that greatly improves the precision and sensitivity of the technique as compared to previous methods \cite{torchinsky2014}. In this setup, data points along a full $360^\circ$ rotation of the scattering plane are rapidly collected and averaged by sweeping $\phi$ at a rate of $\sim 4$ rev/s. The SHG intensity is projected onto a two-dimensional CCD detector, as shown in Fig.~\ref{figure0}(c), with each angle $\phi$ corresponding to a unique fixed position along a circle on the CCD. Further details concerning the setup can be found in Ref.~\citenum{harter2015}. The fundamental beam was generated with a Ti:sapphire regenerative amplifier that produces 60 fs optical pulses, broadened to 150 fs at the sample by dispersion, operating at a 10 kHz repetition rate and with an 800 nm center wavelength ($\hbar\omega = 1.5$ eV). The laser fluence at the sample was $\sim 2$ mJ/cm$^2$ and the CCD exposure time was 300 s for each image. After subtraction of a constant background resulting from CCD pixel readout noise, rotational anisotropy curves were extracted from the CCD images by radially integrating over the annular illumination region and then binning the data in 6$^\circ$ intervals. We supplement our RA-SHG measurements with pump-probe transient reflectivity experiments, which were carried out using 70 fs pulses from the same 800 nm laser with a pump fluence of $\sim 4$ mJ/cm$^2$ in a crossed linear polarization geometry.

\section{Experimental Results and Discussion}

During SHG, two photons at frequency $\omega$ are absorbed and one photon at frequency $2\omega$ is emitted in a coherent quantum process between an occupied electronic level and two unoccupied levels. The total and partial density of states calculated by DFT for CaFeAs$_2$ is shown in Fig.~\ref{figure0}(e). Like the other iron-based superconductors, states near $E_F$ have dominantly Fe $3d$ character with some As $4p$ hybridization. Interestingly, the density of states has a large contribution from Ca $3d$ orbitals starting $\sim 2$ eV above $E_F$. Our RA-SHG experiments at $2\hbar\omega = 3$ eV will therefore mainly probe states derived from the Fe and Ca $3d$ orbitals. Figure \ref{figure1}(a-d) show raw CCD images taken for Ca$_{0.73}$La$_{0.27}$FeAs$_{2}$ at a sample temperature of 70 K, where the crystal structure is monoclinic ($T > T_S$). We measure a very strong SHG response. This is consistent with a noncentrosymmetric point group because inversion-symmetric systems forbid bulk electric dipole SHG and only allow weaker electric quadrupole and other higher-order responses \cite{boyd1991}. For comparison, under the same experimental conditions the bulk electric quadrupole response of Sr$_2$IrO$_4$ is $\sim 100$ times weaker \cite{zhao2015}. Based on the diffraction data of Ref.~\citenum{katayama2013}, the atomic positions in the unit cell of Ca$_{1-x}$La$_{x}$FeAs$_{2}$ differ from a centrosymmetric structure by less than 0.024 \AA, with the largest deviations from centrosymmetry occurring in the Ca$_{1-x}$La$_{x}$ planes. This suggests that Ca $3d$-derived states contribute significantly to the signal that we observe, as predicted by our DFT calculations.

To obtain the point group symmetry of the crystal above $T_S$, we extract rotational anisotropy curves from the raw CCD images [Fig.~\ref{figure1}(e-h)] and fit them to the square of Eq.~\ref{eqn1}, taking advantage of the intrinsic permutation symmetry of SHG ($\chi_{ijk} = \chi_{ikj}$) to reduce the number of free parameters. Using a nonlinear optical susceptibility tensor constrained by the noncentrosymmetric point group $C_2$ established from diffraction measurements \cite{katayama2013}, the only nonzero tensor elements are \cite{boyd1991}:
\begin{eqnarray*}
\chi_{xyz} = \chi_{xzy}, \chi_{xxy} = \chi_{xyx}, \chi_{yzx} = \chi_{yxz},\\*
\chi_{zyz} = \chi_{zzy}, \chi_{zxy} = \chi_{zyx}, \chi_{yxx}, \chi_{yyy}, \chi_{yzz},
\end{eqnarray*}
where we define the orthogonal $(x,y,z)$ coordinate system with respect to the crystal axes such that the $y$-axis is parallel to \textbf{b} and the $z$-axis is parallel to \textbf{c}, which is perpendicular to the FeAs planes and the sample surface, as shown in Fig.~\ref{figure0}(a). [We note that the two-fold rotational symmetry in the crystal is about the $b$-axis, \textit{not} the sample surface normal; we therefore do not expect, and do not observe, two-fold rotational symmetry in our RA-SHG patterns.] We use the eight independent tensor elements as free parameters in a single global fit, the result of which is displayed in Fig.~\ref{figure1}(e-h). Excellent agreement with the data is achieved with the following values, normalized so that $\chi_{yzz} = 1$:
\begin{align*}
\chi_{yzz} &\equiv +1.000, &\chi_{zyz} &= -0.841,\\*
\chi_{yxx} &= -0.788, &\chi_{yyy} &= -0.524,\\*
\chi_{xxy} &= -0.434, &\chi_{xyz} &= -0.149,\\*
\chi_{zxy} &= +0.125, &\chi_{yzx} &= +0.066.
\end{align*}
We emphasize that in addition to the rotational anisotropies, the relative intensities of the different polarization geometries are also replicated by the fit. Furthermore, the data are incompatible with an electric dipole response from any point group of higher symmetry, as attempts to fit the data with such symmetries do not show even qualitative agreement with the data. The high quality of the fits demonstrates that our SHG signal originates from the bulk electric dipole response of a $C_2$ point group. This result confirms that the 112-type materials lack a center of inversion with implications for the potential mixing of spin-singlet and spin-triplet Cooper pairs in the superconducting state. We point out that if the sample were to contain small twin structural domains, with As chains aligned at $90^\circ$ with respect to each other, we would expect to measure a superposition of the observed RA-SHG patterns and their $90^\circ$-rotated counterparts. Instead we observe a single set of patterns with fixed alignment across multiple locations on the sample over a length scale of $\sim 0.5$ mm, showing that the sample is macroscopically detwinned and the As zigzag chains are globally aligned.

\begin{figure}[htb]
\includegraphics{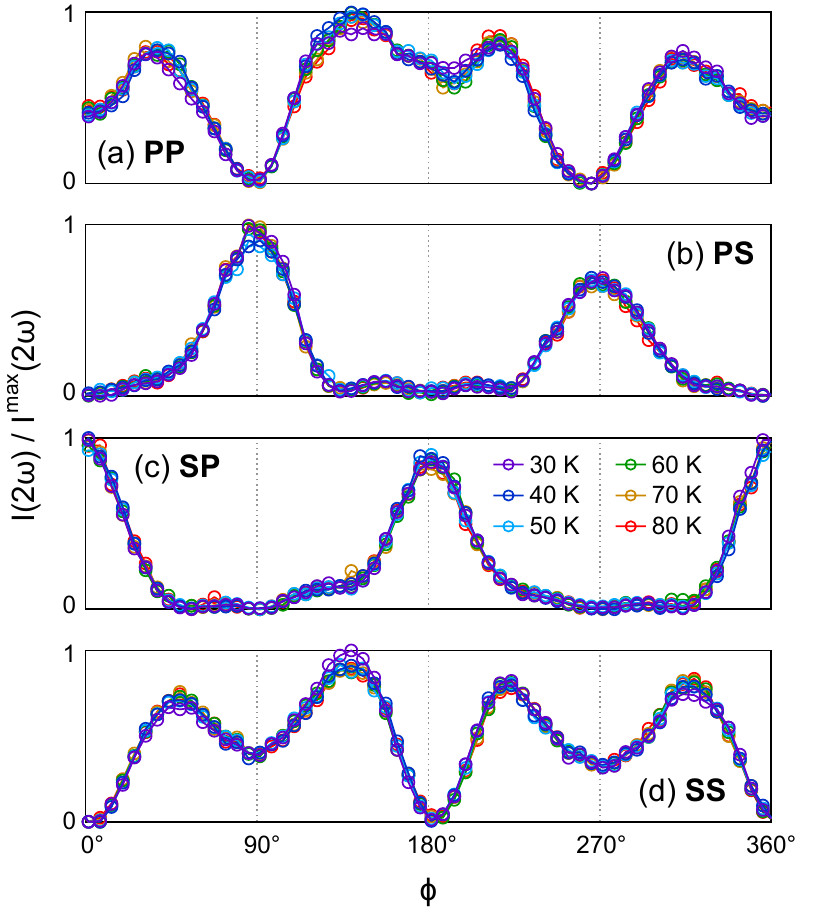}
\caption{\label{figure2} Temperature dependence of RA-SHG for Ca$_{0.73}$La$_{0.27}$FeAs$_{2}$. Rotational anisotropy curves for (a) P$_\mathrm{in}$-P$_\mathrm{out}$, (b) P$_\mathrm{in}$-S$_\mathrm{out}$, (c) S$_\mathrm{in}$-P$_\mathrm{out}$, and (d) S$_\mathrm{in}$-S$_\mathrm{out}$ polarization geometries taken from 30 to 80 K in 10 K steps. Curves are normalized by their respective average values in order to remove the effects of long-term laser power drift. There is no detectable change in anisotropy with temperature across the structural phase transition.}
\end{figure}

We can measure possible changes in point group symmetry across the reported monoclinic-to-triclinic structural transition at $T_S = 58$ K by examining the temperature dependence of the SHG rotational anisotropy. The two possible point groups for a triclinic lattice are $C_i$ and $C_1$. The former case has no bulk electric dipole SHG response because of the presence of inversion symmetry, which would lead to a dramatic decrease in SHG intensity. The latter case possesses no point group symmetry operations other than the identity, making all nonlinear optical susceptibility tensor elements $\chi_{ijk}$ independent and nonzero (with intrinsic permutation symmetry still requisite). Consequently, we write
\begin{equation}
\chi(T) = \chi^\mathrm{mono} + \Delta\chi(T)
\end{equation}
and associate all changes in the susceptibility tensor at the transition with the temperature-dependent $\Delta\chi(T)$ term, which is expected to turn on below $T_S$ and encodes changes away from the high-temperature susceptibility $\chi^\mathrm{mono}$ induced by the lowering of symmetry in the triclinic phase. The possibility of accurately measuring such a parameter is enabled by our RA-SHG setup, which averages over long time scale laser fluctuations that limit the sensitivity of conventional setups.

\begin{figure}[htb]
\includegraphics{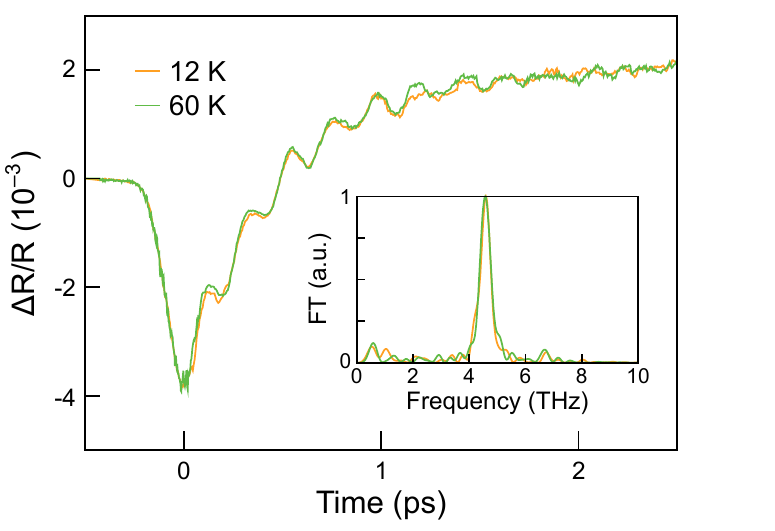}
\caption{\label{figure3} Pump-probe transient reflectivity measurements of Ca$_{0.73}$La$_{0.27}$FeAs$_{2}$, showing the time-resolved relative change in reflectivity after the pump pulse above and below the structural phase transition. Inset shows the corresponding Fourier transform (FT) of the reflectivity after subtracting a long timescale exponential response, as discussed in the text. A phonon resonance at 4.6 THz is present in both measurements.}
\end{figure}

A comparison of rotational anisotropy curves as a function of temperature through $T_S$ is shown in Fig.~\ref{figure2}. We do not observe any detectable change in the symmetry of the crystal to within our experimental resolution [$\Delta I(2\omega)/I(2\omega) < 5$\%], placing bounds on the magnitude of $\Delta\chi(T)$ for 800 nm incident light. Despite the absence of a detectable change in the nonlinear optical susceptibility, we can conclude that the low-temperature point group is $C_1$ since we do not see a suppression of SHG intensity expected if the lattice were to acquire inversion symmetry ($C_i$). The spin order that sets in at $T_N$ should in principle also contribute to a change in the optical susceptibility. Such a change, however, is likely to be overwhelmed by the strong nonmagnetic contribution already present above $T_N$. Indeed, in practice nonlinear optical studies of magnetic structure predominantly focus on centrosymmetric materials where the nonmagnetic bulk electric dipole response is zero \cite{fiebig2004}. The absence of an observable change in the SHG rotational anisotropy suggests that structural changes at the monoclinic-to-triclinic transition are subtle and do not significantly affect the electronic states near the Fermi level that are probed by SHG.

In addition to RA-SHG measurements, which we have shown mainly probe the Fe and Ca $3d$-derived states in the crystal, we also present pump-probe transient reflectivity experiments because such measurements have previously been shown to be sensitive to Raman active phonons involving As atoms in the iron-based superconductors \cite{mansart2009}.  Our results are displayed in Fig.~\ref{figure3}. Immediately after the pump pulse, we observe a drop in reflectivity of order $\Delta R/R \approx 4\times10^{-3}$ followed by a ps-timescale recovery with high-frequency oscillations superimposed. The long timescale behavior is likely due to the relaxation of hot electrons in the system by exchange of energy with the lattice. The high-frequency oscillations, on the other hand, are the result of the coherent excitation of an optical phonon mode.

To isolate the oscillatory component of the response, we fit the reflectivity for $t > 0$ to a constant plus a decaying exponential (decay time $\tau \approx 0.5$ ps) and subtract it from the data. A Fourier transform of the remaining oscillations produces a frequency power spectrum showing an isolated peak at 4.6 THz (inset of Fig.~\ref{figure3}). This frequency corresponds to a phonon energy of 19 meV. To date, we are not aware of any Raman spectroscopy studies of the 112-type materials. The other iron-based superconductor families, however, show Raman active phonons near this energy, including the out-of-plane As $A_{1g}$ mode at 20 to 25 meV reported in the 111-, 1111-, and 122-type materials \cite{zhao2009,zhang2010,um2012,antal2012,teng2012}. A comparable time-resolved reflectivity study using an 800 nm pump reported similar oscillations in Ba(Fe$_{1-x}$Co$_x$)$_2$As$_2$ that they also attributed to the As $A_{1g}$ mode \cite{mansart2009}, suggesting the mode as a likely candidate for the reflectivity oscillations observed here. If this is the case, these vibrational modes of the FeAs planes in Ca$_{1-x}$La$_{x}$FeAs$_{2}$ are comparable to those measured in the other iron-based superconductors despite their differing crystallographic symmetries. Within our experimental resolution, we do not observe a change in the phonon frequency at the monoclinic-to-triclinic structural transition. This observation suggests that the potential landscape of the As atoms does not significantly change across $T_S$ and may indicate that the unit cell distortion at the transition is primarily within the $(a,b)$-plane [as illustrated in Fig.~\ref{figure0}(b)] since the phonon mode that we observe involves strictly out-of-plane motion. Finally, we note that our detection of only the $A_{1g}$ phonon mode does not rule out the possible existence of other Raman-active modes in the frequency range accessible by our experiment. Our results are consistent with Ref.~\citenum{mansart2009}, where a similar transient reflectivity experiment on the 122-type materials also revealed only the $A_{1g}$ mode. They suggested that the phonon excitation mechanism was not impulsive stimulated Raman scattering (in which all optical phonon modes may be excited), but was instead either displacive excitation of coherent phonons or electronic temperature gradient. Either of those mechanisms would selectively excite only the totally symmetric mode.

\section{Summary and Conclusions}

In conclusion, we have used optical second harmonic generation and transient reflectivity experiments to study the structural symmetry of the 112-type iron-based superconductor parent compound Ca$_{0.73}$La$_{0.27}$FeAs$_{2}$. Our RA-SHG results, sensitive to the Fe and Ca $3d$-derived states near $E_F$, allow us to unambiguously identify the crystallographic point group as $C_2$ in the monoclinic phase and $C_1$ in the triclinic phase. The strong electric dipole nonlinear optical response highlights the noncentrosymmetric nature of the crystal, and the observed alignment of the $a$ and $b$ axes at multiple locations implies intrinsically detwinned samples. Pump-probe transient reflectivity measurements uncover a Raman active phonon at 4.6 THz associated with the out-of-plane motion of As atoms. We observe no changes in RA-SHG or transient reflectivity across the monoclinic-to-triclinic structural transition, showing that it does not significantly affect electronic states near the Fermi level and possibly suggesting a weak coupling between the lattice and magnetic order. These subtle but compelling properties make the 112-type compounds a unique and important addition to the iron-based superconductor family of materials.

\section*{ACKNOWLEDGMENTS}

RA-SHG and transient reflectivity experiments were supported by the U. S. Department of Energy under award number DE-SC0010533. Instrumentation for the RA-SHG setup was partially supported by a U. S. Army Research Office DURIP award under grant number W911NF-13-1-0293 and the Alfred P. Sloan Foundation under grant number FG-BR2014-027. D.H. also acknowledges funding from the Institute for Quantum Information and Matter, an NSF Physics Frontiers Center (PHY-1125565) with support of the Gordon and Betty Moore Foundation through grant number GBMF1250. Work at UCLA was supported by the U.S. Department of Energy Office of Basic Energy Sciences under award number DE-SC0011978.

\end{document}